# Selected List of Low Energy Beam Transport Facilities for Light-Ion, High-Intensity Accelerators

L.R. Prost[#], Fermilab[*], Batavia, IL 60510, USA

***Disclaimer***: *This document does not pretend listing all the LEBTs that have been built, operated or designed around the world. It merely gives an overview picture of the technological choices made for light-ion, high-intensity LEBTs, which appeared to be relevant if one would want some background information in the process of designing a similar beam line.*

## 1  Introduction

The Proton Improvement Plan II (PIP-II) at Fermilab is a program of upgrades to the injection complex[1]. At its core is the design and construction of an 800-MeV, 2-mA $H^-$ CW superconducting linac.

To validate various concepts for the front-end of such machine (i.e. first ~30 MeV), a test accelerator (a.k.a. PXIE) is under construction[2]. During the design effort, it was useful to look at other facilities to identify peculiarities that the PXIE beam might present. In this instance, we concern ourselves with the design schemes of Low Energy Beam Transport lines (LEBT) only.

PXIE includes a 2 m-long LEBT, which takes an up to 10 mA DC, 30 keV $H^-$ beam from the ion source to a Radio-Frequency Quadrupole (RFQ). In the process of searching for the relevant information, it became apparent that merely having a list of machines with similar beam parameters for the front-end would be valuable even before going into the deeper details of each individual design.

## 2  List of facilities

A survey of some of past, present and future light-ion high-intensity LEBTs around the world, is summarized in Table 1. It includes LEBTs, which are or have been operational as well as some that are or remained at the design stage.

Table 1: List of light ion, high intensity accelerator facilities, decommissioned, operating or proposed.

| Institution | Accelerator | Ion species | $I_{Beam}$, mA | $E_{Beam}$, keV (or keV/u) | Gen. perv. ×10⁻³ (Eq.1) | Pulse length, ms | Pulse frequency, Hz | LEBT design choice | Operating/ Design parameters |
|---|---|---|---|---|---|---|---|---|---|
| RAL | FETS[4] | H⁻ | 60 | 35 | 5.9 | 2 | 50 | 3 solenoids | Operating[*] |
| CERN | LINAC 4[5,6] | H⁻ | 80 | 45 | 5.4 | 0.4 | 2 | 2 solenoids | Design[†] |
| SSC | SSC injector[7,8,9] | H⁻ | 30 | 35 | 3.0 | 0.007-0.035 | 10 | Electrostatic[#] | Design |
| LANL | GTA[10] | H⁻ | 50 | 35 | 5.0 | 2 | 5 | 2 solenoids | Operating[‡] |
| LBNL | AGS pre-injector[11] | H⁻ | 70 | 35 | 6.9 | 0.45 | 5 | 2 solenoids | Operating |
| DESY | HERA-Linac3 pre-injector[12] | H⁻ | 20 | 18 | 5.4 | 0.035 | <1 | 2 solenoids | Operating |
| ESS/INFN-LNS | ESS[13,14] | H⁺ | 90 | 75 | 2.9 | 2.86 | 14 | 2 solenoids | Design |
| GSI | FAIR proton Linac/SILHI[15,16] | H⁺ | 100 | 95 | 2.2 | 0.036 | 4 | 2 solenoids | Design |
| KEK/JAEA | J-PARC[17,18] | H⁻ | 55 | 50 | 3.2 | 0.5 | 25 | 2 solenoids | Operating |
| ORNL | SNS[19,20] | H⁻ | 50 | 65 | 2.0 | 1 | 60 | 2 einzel lenses | Operating |
| IHEP | CSNS[21] | H⁻ | 50 | 50 | 2.9 | 500 | 25 | 3 solenoids | Operating |
| FNAL | Linac[22] | H⁻ | 60 | 35 | 5.9 | 0.145 | 15 | 2 solenoids | Operating |
| INFN-LNL | SPES/TRIPS[23,24] | H⁺ | 35 | 80 | 1.0 | DC | | 2 solenoids | Operating |
| Soreq NRC | SARAF[25,26] | H⁺, D⁺ | 5 | 20 | 1.1 [$] | DC | | 3 solenoids | Operating |
| SCK-CEN | MYRRHA/MAX[27,28] | H⁺ | 5 | 30 | 0.6 | DC | | 2 solenoids[&] | Design |
| KAERI | KOMAC[29] | H⁺/H⁻ | 30/3 | 50 | 1.7 [$] | DC | | 2 solenoids | Operating |
| LANL | LEDA[30] | H⁺ | 110 | 75 | 3.5 | DC | | 2 solenoids | Operating |
| BARC | LEHIPA[31,32] | H⁺ | 30 | 50 | 1.7 | DC | | 2 solenoids | Design |
| IMP | C-ADS Injectors[33,34,35] | H⁺ | 30 | 35 | 3.0 | DC | | 2 solenoids | Operating[††] |
| FNAL | PXIE[2] | H⁻ | 10[**] | 30 | 1.2 | DC | | 3 solenoids | Operating |
| CEA-Saclay | IFMIF/EVEDA[36,37] | D⁺ | 140 | 50 | 4.1 | DC | | 2 solenoids | Design[##] |
| LBNL | ESQ injector[38] | H⁻ | 45 | 100 | 0.9 | DC | | Electrostatic quadrupoles | Operating |
| Goethe University | FRANZ[39,40] | H⁺ | 50 | 120 | 0.8 | DC | | 4 solenoids | Design[‡‡] |
| LANL | LANSCE[41] | H⁺ | 35 | 35 | 3.5 | DC | | 2 solenoids | Design[⁂] |
| TRIUMF/AES | CDS[42,43] | H⁻ | 10 | 40 | 0.8 | DC | | 1 solenoid | Operating |
| GANIL | SPIRAL 2[44] | D⁺ | 6.5 | 20 | 0.7 | DC | | 2 solenoids + quadruplet quads | Design[£] |

[*] at 25 Hz only
[†] parameters demonstrated on a test stand, but not concurrently
[#] normal quadrupoles, helical quadrupoles and einzel lenses were considered. LEBTs with einzel lenses and helical quadrupoles were tested but not to full beam design parameters.
[‡] design was 2% duty factor (measurements only up to 1% duty factor)
[&] evolution from EURISOL injector design, which included 4 solenoids
[$] for H⁺
[††] at 10 mA only
[**] max current; 5 mA nominal
[##] achieved 120 mA DC with H⁺ and 145 mA with D⁺ @ 9.5% duty factor only
[‡‡] commissioning activities with 3.5 mA, 14 keV He⁺, which corresponds to a generalized perveance of 2.7×10⁻³
[⁂] replacement of the existing Cockroft-Walton based injectors
[£] nominal parameters demonstrated in pulsed mode

## 3 Discussion

For the range of beam currents (5-140 mA) and energies (20-120 keV) listed here, most designs employ a set of solenoids to transport the beam from the ion source to the next accelerating structure, and it is independent of the beam time structure (i.e. from short pulses to DC). These 'magnetic' LEBTs operate in a regime where the beam is highly neutralized, reducing significantly the potential for emittance growth due to the space charge. Yet, if the beam is pulsed, it gets significantly mismatched during the Space Charge Compensation (SCC) transient, which is often addressed with an additional fast chopping system downstream and/or some limiting apertures.

Very few 'electrostatic' LEBTs were found either operational or designed (only 3 identified in the table). They operate in a fully un-neutralized mode by default since the neutralizing particles are swept away by the large electric field of the focusing elements. Thus, it eliminates the difficulties related to the transient time associated with SCC. On the other hand, space charge often limits the maximum transportable beam current and space charge non-linearities may become problematic.

A discussion of pros and cons for both types of LEBTs can be found in Ref. [3]. Below, we suggest using the beam generalized perveance as a figure of merit for comparing the potential for space charge-induced emittance growth between the facilities listed in Table 1 in case of incomplete neutralization.

### 3.1 Beam perveance and potential for emittance growth

The beam current and energy of the facilities listed in Table 1 vary over a fairly wide range of values. Since the magnitude of space charge forces play an important role in the design of the beam line optics, in order to help in making comparisons between the different systems, included in Table 1 is the unit-less parameter called the generalized perveance defined as[45]:

$$K = \frac{I_{Beam}}{V_{ext}^{3/2}} \cdot \left[\frac{1}{4\pi\varepsilon_0 (2q/m)^{1/2}}\right] \quad (1)$$

where $I_{Beam}$ is the beam current in Amps, $V_{ext}$ is the ion source extraction voltage in Volts, $q$ is the total electrical charge in Coulombs, $m$ is the mass of the particle in kg and $\varepsilon_0$ is the permittivity of vacuum in SI units. Thus, for all the machines listed in Table 1, the beam perveance lies between $6\times10^{-4}$ and $7\times10^{-3}$. Generally speaking, the beam perveance sets a practical limit for un-neutralized transport where lumped focusing is no longer possible because the space charge forces cause the beam to diverge too fast between consecutive lenses. A practical estimate of what this limit may be, can be derived (e.g.: following Ref. [46] Section 3) and is found to be ~$1.2 \times 10^{-2}$, almost an order of magnitude above any of the values listed in Table 1 for which un-neutralized transport was realized, the largest being $2\times10^{-3}$ for Oak Ridge National Laboratory's Neutron Spallation Source.

The practical limitation associated with a large beam perveance just mentioned is not necessarily what makes the realization of space charge dominated transport often unattractive. Instead, this is the potential for emittance growth associated with large tune depression concurrent with the imperfections of a real beam and machine even in the linear regime. An analytical formulation for the simplest model of continuous focusing and paraxial approximation is given and discussed in some detail in Ref. [47]. While there are other components that determine the evolution of the emittance in any accelerator, the predominance of LEBTs employing a neutralization scheme shows the importance given to the possibility of emittance growth due to space charge in the design of a beam transport line.

### 3.2 Number of focusing elements

There are many details that go into the design of a LEBT, well beyond the discussion of the preceding section, which merely illustrates how some basic parameters (such as beam energy and beam current) may set some limits to what is theoretically and practically achievable. On the other hand, Table 1 clearly indicates that solenoid focusing has been and still is the most common choice for LEBTs around the world (with the most notable exception of the SNS front-end). From a theoretical point of view, only 2 free parameters are necessary to match the Twiss functions ($\alpha$, $\beta$) of the beam produced by an ion source to those required at the entrance of a RFQ or another accelerating structure. For instance, in a single solenoid configuration, these 2 parameters are the solenoid current (i.e. focusing strength) and the solenoid position, given that the beam line geometry and the ion source parameters are fixed. That was the choice made by TRIUMF/AES for their Contraband Detection System (CDS) injector. However, the most common choice for a magnetic LEBT is to use 2 solenoids in order to be able to accommodate various tunes of the ion source.

Then, the choice of adding more focusing elements appears to be mainly dictated by auxiliary considerations often associated with other requirements for the LEBT. For instance, the beam line might include a bend and/or a chopping system, which in turn might require the beam to have specific parameters at its entrance. This is often accommodated with one more solenoid (for a total of 3), but for example, for FRANZ at the Goethe University in Frankfurt, it was decided to have 2 solenoids before and 2 solenoids after a complex E×B chopping system. For PXIE, a 3-solenoid LEBT has been chosen to accommodate a bending magnet and a robust chopper. Adding more focusing elements might also be required when the LEBT has to accommodate multiple ion species in a common beam line (e.g.: SPIRAL 2 at GANIL, which employs quadruplet quads in addition to solenoids)

### 3.3 PXIE LEBT

The PXIE LEBT design choices are discussed in some detail in Ref [46]. One of its peculiarities is the fact that the beam transport is space charge dominated, by design, over the last ~1m before the RFQ. The PXIE beam perveance is one of the lowest listed in Table 1, making it relatively easy to focus even without any neutralization. Also, the PXIE LEBT transport scheme assumes that the transition between neutralized transport and space charge dominated transport occurs where the beam transverse distribution is close to being uniform, hence limiting the potential for emittance growth due to relaxation into a thermal equilibrium. Scraping of the tail particles to limit non-linear space charge forces is also part of the scheme.

## 4 Conclusion

A list of LEBTs for light ion, high intensity accelerator facilities has been compiled. Most designs employ a transport scheme with at least 2 solenoids and high degrees of beam neutralization, although details of the beam lines may vary significantly. While solely based on the value of the generalized perveance space charge-dominated transport is not precluded, the overwhelming use of a neutralized transport scheme suggests that the potential for emittance growth due to relatively large tune depressions concurrent with non-ideal, non-stationary particle distributions and phase-space distortions due to space charge non-linearities are prime considerations in the design of Low Energy Beam Transport lines.

## 5 Acknowledgement

The author is grateful to A. Shemyakin for his critiques and guidance, as well as for his suggestion to write such a document.